# Factors controlling the energetics of the oxygen reduction reaction on the Pd-Co electro-catalysts: Insight from first principles.

Sebastian Zuluaga and Sergey Stolbov

Physics Department University of Central Florida, Orlando, FL

## Abstract

We report here results of our density functional theory based computational studies of the electronic structure of the Pd-Co alloy electrocatalysts and energetics of the oxygen reduction reaction (ORR) on their surfaces. The calculations have been performed for the (111) surfaces of pure Pd, $Pd_{0.75}Co_{0.25}$ and $Pd_{0.5}Co_{0.5}$ alloys, as well as of the surface segregated $Pd/Pd_{0.75}Co_{0.25}$ alloy. We find the hybridization of $d$Pd and $d$Co electronic states to be the main factor controlling the electrocatalytic properties of $Pd/Pd_{0.75}Co_{0.25}$. Namely the $d$Pd – $d$Co hybridization causes low energy shift of the surface Pd $d$-band with respect to that for Pd(111). This shift weakens chemical bonds between the ORR intermediates and the $Pd/Pd_{0.75}Co_{0.25}$ surface, which is favorable for the reaction. Non-segregated $Pd_{0.75}Co_{0.25}$ and $Pd_{0.5}Co_{0.5}$ surfaces are found to be too reactive for ORR due to bonding of the intermediates to the surface Co atoms. Analysis of the ORR free energy diagrams, built for the Pd and $Pd/Pd_{0.75}Co_{0.25}$, shows that the co-adsorption of the ORR intermediates and water changes the ORR energetics significantly and makes ORR more favorable. We find the onset ORR potential estimated for the configurations with the O – OH and OH – OH co-adsorption to be in very good agreement with experiment. The relevance of this finding to the real reaction environment is discussed.

# 1. Introduction

Fuel cells, such as the proton exchange membrane fuel cells (PEMFC) and the direct methanol fuel cells (DMFC), convert the hydrogen chemical energy into electric power. As clean renewable sources of energy they can offer great advantages for various applications, however, a number of obstacles remain to their large scale implementation. First, fuel cells are unacceptably expensive. Since the Pt-based catalysts, used in both electrodes of the fuel cell, make up a major part of the cost, search for new electrocatalytic materials with a reduced loading of precious metals is critical for commercialization of PEMFC and DMFC. Second, performance of both PEMFC and DMFC suffers from low rate of the oxygen reduction reaction (ORR) on the Pt cathode, which decreases the onset potential for ORR (~0.9V vs standard hydrogen electrode (SHE), compared to 1.23 V (SHE) of ideal potential), and hence reduces fuel cell efficiency [1].

Clearly, the great advantages of fuel cells can be efficiently utilized only if the cost of the electrodes is dramatically reduced and their electrocatalytic properties are significantly improved. The search of new electrocatalysts for ORR is conducted in several directions. One of them focuses on the systems including the Pt monolayer deposited on a Pt-free substrate. The Adzic and Mavrikakis research groups combining experimental studies and first principles calculations have made a significant progress in this direction [2-6]. The authors have found some Pt/M structures with the ORR activity comparable or even higher than that on bulk Pt. These works also provide insight into the mechanisms of formation of the layered surface structures and their effect on the ORR energetics. Since such catalysts are mostly synthesized in the form of the 2 - 5 nm nanoparticles, surface atoms (Pt) make a significant fraction of volume of the nanoparticles. The Pt load is still high, which is a disadvantage of these systems. It is thus not surprising that much effort has been made to find efficient Pt-free electrocatalysts for ORR. Along this direction, Pd alloys are extensively studied as promising materials. In particular, $Pd_{1-x}M_x$ alloys where M=Fe, Co, Cu have shown enhanced rate of ORR [7-12].

In this work we focus on the $Pd_{1-x}Co_x$ system. These alloys exhibit activity toward ORR comparable to Pt one [7, 12], highly tolerant to the presence of methanol [8,13], which is important for use in DMFC, and they are much less expensive that the Pt-based catalysts. The best electrocatalytic performance of these alloys has been observed for the Co concentration x = 0.2 – 0.33 [8,12]. For such concentration range the alloys still maintain the fcc structure. At elevated temperature they undergo the surface segregation, which lead to formation the Pd monolayer (skin) on the alloy surface [13]. However, as suggested in Ref. 13 and confirmed in the present work, this segregation is a wanted effect, which leads to the enhancement of the ORR rate.

Naturally, understanding of the mechanisms underlying such promising properties of these materials is of high interest now, and some effort has been made to gain insight into these mechanisms from first principles calculations within the density functional theory. Suo et al. [14] modeled Pd-Co alloy surfaces by a three layer slab. They have calculated the energies of oxygen atomic adsorption for some configurations of the Pd and Co atoms in this simple supercell. The results have brought the authors to the conclusion that the Pd lattice strain effects, caused by Co alloying, are responsible for the enhancement of the ORR rate in the system. Lamos and Balbuena used a three layer slab made up with Pd, $Pd_{0.5}Co_{0.5}$, and $Pd_{0.75}Co_{0.25}$ layers to model the segregated $Pd_{0.75}Co_{0.25}$ (111) surface [15]. The authors have calculated the adsorption energies of intermediates and build the free energy reaction diagrams for two possible reaction pathways using the Nørskov and co-worker model [16].

These works provide some insight into the nature of electrocatalytic activity of the Pd-Co alloys, however, for deep understanding of the alloying effect on the ORR rate, many key questions have still to be answered. For example, the authors of Ref. 14 argue that contraction of Pd – Pd bonds upon alloying is responsible for the enhanced ORR rate based on the change in the oxygen binding energy ($E_B(O)$) by 0.05 eV upon the contraction. However, this change is too small to cause a significant effect on the properties in question. On the other hand, the Pd atoms in the overlayer make bonds with Co atoms located in the second layer. Cobalt has quite delocalized $d$-states, which have to result in strong hybridization with the $d$Pd-states. The spin-polarized local densities of electronic states (LDOS) should form a wide $d$-band overlapping with the $d$Pd-states. One can thus expect a significant modification of the Pd $d$-band upon the $d$-Co – $d$Pd hybridization, which, in turn, will change the adsorption energies. This important effect has not been studied yet for this system.

Next, the calculations in Ref. 14 have been performed for the adsorbate coverage of 0.25 monolayer (ML). Meanwhile, as shown for other materials, higher coverage, co-adsorption of different reaction intermediates and water can significantly change the reaction energetics [16 - 18]. To our knowledge, these effects have not been analyzed so far for the Pd – Co alloys.

ORR is a multi-step reaction with complex energetics, and it is not surprising that significant effort has been made in finding a simple descriptor for preliminary screening materials in the course of design of efficient electrocatalysts. Authors of Ref. 16 and 19 have built the $E_B(O) - E_B(OH)$ diagrams (hyper-volcano diagrams) for a number of transition and noble metals, which show, for example, that an optimal catalyst for ORR has to have lower $E_B(O)$ and $E_B(OH)$ values, than those for Pd. Furthermore, it has been shown that for some transition and noble metal surfaces there are linear relations between the O and OH, as well as between O and OOH binding energies [20], which suggest that $E_B(O)$ can be used as a descriptor of the ORR activity [21, 22]. It is important to learn whether these relations are in effect for higher coverage and co-adsorption the ORR intermediate and water, and whether trends in change in $E_B(O)$ for low coverage configurations correlate with the ORR energetics for the cases of high coverage and co-adsorption.

Finally, since the nature of the effect of the Pd overlayer in the segregated structures on the ORR energetics is not quite clear, it is useful to calculate and compare the properties of the Pd-Co alloys with and without the overlayer.

In the present work, we address the issues raised above. We report here results of accurate systematic computational studies of various factors which may control the ORR rate, including effects of co-adsorbed intermediates and water on the ORR energetics. We compare and contrast the results obtained for Pd – Co alloys with varying Co concentration x = 0, 0.25, and 0.5, as well as the $Pd_{0.75}Co_{0.25}(111)$ alloy covered with one monolayer of Pd ($Pd/Pd_{0.75}Co_{0.25}$). The reaction energetics is traced to the electronic structure of the alloy surfaces in order to reveal the main factors controlling the ORR rate in the system.

## 2. Model

The ORR is a complex multi-electron reaction that may include many steps. As summarized by Adsic [23] two main pathways are possible: a) direct four electron reduction to $H_2O$ (in acid media):

$$O_2 + 4H^+ + 4e^-(U) \rightarrow 2H_2O, \quad (1)$$

b) peroxide pathway:

$$O_2 + 2H^+ + 2e^-(U) \rightarrow H_2O_2, \quad (2)$$

followed by

$$H_2O_2 + 2H^+ + 2e^-(U) \rightarrow 2H_2O \quad (3)$$

The second step in the pathway (b) has a very high reversible potential that significantly reduce efficiency of ORR.

ORR has been extensively studied on Pt surfaces. It has been suggested the 4-electron pathway is predominant for Pt [23]. The DFT based calculations show that the activation energy barriers for $O_2$ dissociation on flat Pt surfaces are high and the molecular adsorption is thus preferred [24]. In this case ORR may proceed through the following steps:

$$O_2 + * \rightarrow O_2^* \quad (4)$$

$$O_2^* + (H^+ + e^-) \rightarrow HO_2^* \quad (5)$$

$$HO_2^* + (H^+ + e^-) \rightarrow H_2O + O^* \quad (6)$$

$$O^* + (H^+ + e^-) \rightarrow HO^* \quad (7)$$

$$HO^* + (H^+ + e^-) \rightarrow H_2O + * \quad (8)$$

In this notation, "*" denotes the adsorption site at the cathode surface. If oxygen dissociate upon adsorption, the ORR pathway is as follow:

$$O^* + H^+ + e^- \rightarrow HO^* \quad (9)$$
$$HO^* + H^+ + e^- \rightarrow H_2O + * \quad (10)$$

The electrocatalytic kinetics is a complicated phenomenon, which involves the electron and proton transfers between two electrodes with different Fermi-levels. It has been shown, however, that some approaches from conventional heterogeneous catalysis and gas phase reactions can be used to describe this phenomenon by choosing a reference which links gas-phase and electrocatalytic quantities. In this work we use the technique proposed by Nørskov and co-authors [16], in which they set the reference potential as $\mu(H^+ + e^-) = \frac{1}{2}\mu(H_2)$ that allows replacing the energy of the $n(H^+ + e^-)$ transfer with the energy of hydrogen molecule with a corresponding multiplier. Within this approach, the reaction free energy $\Delta G$ is calculated for each reaction step, which is defined as the difference between free energies of the initial and final states of the step. In general, it includes six terms:

$$\Delta G = \Delta E + \Delta ZPE - T\Delta S + \Delta G_U + \Delta G_{field} + \Delta G_{pH}. \qquad (11)$$

Here $\Delta E$ is the reaction energies obtained from DFT total energies of the catalyst surface with possible configurations of adsorbed reactants or intermediates. The DFT-based calculations of vibrational frequencies of adsorbates are used to define zero point energy corrections $\Delta ZPE$. Entropic contribution $T\Delta S$ is approximated by the gas phase reaction entropy of reactants or intermediates taken from a NIST database [25] (translational contributions are subtracted for adsorbed species). If a reaction step involves the electron and proton transfer, the relevant bias effects are taking into account by shifting the energy by $\Delta G_U = -eU$, where $U$ is the electrode potential and $e$ is a transferred charge. The term $\Delta G_{field}$ is a contribution of interaction of an adsorbate with local electric field in the electric double layer formed in the vicinity of cathode [26]. Finally, for non-zero pH the concentrational entropy correction is added: $\Delta G_{pH}(pH) = kT*ln10*pH$. As $\Delta G$ is calculated for each step of electrocatalytic reaction, a diagram of the free energies of these configurations is built as a function of the electrode potential.

Within this model, the onset potential $U_0$ can be estimated as the maximum value of $U$ at which the reaction is still exothermic. If $G(O) > 2G(OH)$, this condition is achieved for $U_0 = G(OH)$, otherwise, $U_0 = G(O) - G(OH)$. Here $G(O)$ and $G(OH)$ denote the free energies of the $H_2 + O^*$ and $1/2H_2 + OH^*$ states, counted from the free energy of the final state ($H_2O + *$).

The Eq. (11) includes various terms and it is useful to select the main factors changing the energetics of the diagram upon variation of the material content and/or the surface morphology. Importantly, it has been shown [16] that the onset potential ($U_o$) of the catalysts with molecular adsorption is determined by steps (7) and (8) which are the same as the steps (9) and (10) for dissociative adsorption. The changes in free energies of the $O^* + 2(H^+ + e^-)$ and $HO^* + (H^+ + e^-)$ states are thus the key characteristics of electro-catalysts for both molecular and dissociative adsorption of $O_2$. Taking as a reference the free energy of the final state of the reaction ($H_2O$ in gas phase), one can express $\Delta E$ in Eq. (11) through binding energies of the intermediates and total energies of molecules in gas phase. For the states, described by Eqs. 9 and 10, it makes:

$\Delta E(O) = E_{tot}(H_2) + E_{tot}(O) - E_{tot}(H_2O) - E_B(O^*) \qquad (12)$

$\Delta E(OH) = \frac{1}{2}E_{tot}(H_2) + E_{tot}(OH) - E_{tot}(H_2O) - E_B(OH^*) \qquad (13)$

The first three terms in the right sides of Eqs. (12) and (13) represent the total energy of molecules in gas phase which do not depend on the catalyst material properties. It is thus found that contribution of the catalyst to $\Delta E$ is totally determined by the binding energies of O and OH. The $\Delta ZPE$ and $T\Delta S$ are determined by vibration frequencies of the molecules which just slightly change upon adsorption. One can thus conclude that the energetics of the reaction free energy phase diagram is essentially determined by binding energies of the O* and OH* intermediates.

## 3. Computational Details

The experimentally observed Pd-Co electro-catalysts are found to be in form of 4 nm to 11 nm nanoparticles. [12, 27]. Particles of such a size range have large flat facets developed at their surfaces. We thus use the flat surface approximation to describe the catalytic properties of this system. Since the materials under consideration have the fcc structure, we calculate the ORR characteristics on Pd-Co(111) surface. This surface is known to be the most stable one and we expect that the (111) facets to be dominating for these particles.

For all system under consideration, the electronic structure, energetics and equilibrium atomic configurations are obtained using the VASP5.2 code [28] with projector augmented wave potentials [29] and the Perdew-Burke-Ernzerhof (PBE) version of the generalized gradient approximation (GGA) for the exchange and correlation functional [30]. All systems, except for clean Pd, were calculated taking into account spin polarization. In order to maintain periodicity we use supercells with a 5 layer Pd-Co slab and vacuum layer of 15 Å. For all calculations the supercells had the (2x2) in-plane periodicity (see Fig. 1). The (7x7x1) $k$-point samplings in Brillouin zone used in this work provide sufficient accuracy for the characteristics obtained by integration in the reciprocal space. The cut of energy of 400 eV was used for the plane wave expansion of wave functions and the 600 eV cut of energy was used for the charge density. To achieve structural relaxation, a self-consistent electronic structure calculation was followed by calculation of the forces acting on each atom. Based on this information the atomic positions were optimized to obtain equilibrium geometric structures in which forces acting on atoms do not exceed 0.02 eV/Å.

To characterize strength of bonding of intermediates (Int = O, OH) on the catalyst surface we used the adsorption energy defined as follows:

$$E_B(Int) = E_{tot}(slab) + E_{tot}(Int) - E_{tot}(Int/slab), \qquad (14)$$

where the three $E_{tot}$ terms denote the total energies per supercell calculated for the O or OH adsorbed on the surface, clean Pd-Co slab, and isolated O atom or OH, respectively. Given the total energies of stable systems are negative, $E_B(Int)$ is positive if adsorption of a specie on the slab is favorable.

To obtain $\Delta ZPE$ used in Eq. 11, we have calculated the vibrational frequencies of the adsorbed O and OH using the finite-difference method. Since masses of the ORR intermediates are much smaller than those of the substrate, only the adsorbate modes were taking into account with the frozen slab atoms. Five displacements were used for each direction with the step of 0.015 Å. The zero point energies obtained from the vibrational frequencies were used to calculate the *ΔZPE* contributions to the reaction free energies. Entropic contributions to the reaction free energies were calculated as described in Section II.

The geometric structures of clean and adsorbed surfaces shown in this article have been plotted using the Xcrysden software [31].

## 4. Results and Discussion

In this work we have studied the electro-catalytic properties of four systems: Pd(111), $Pd_{0.75}Co_{0.25}$(111), $Pd_{0.5}Co_{0.5}$(111) and $Pd_{0.75}Co_{0.25}$(111) alloy covered with one monolayer of Pd (Pd/$Pd_{0.75}Co_{0.25}$), to model the surface segregation mentioned in Introduction. First the calculations were performed for the clean surfaces, the structural relaxation has been achieved and the total energies obtained for all above alloys. Next the adsorption energies were calculated for various configurations of O, OH and $H_2O$ on the surfaces.

### 4.1. O and OH adsorption with 0.25 ML coverage

First, for all systems under consideration, the binding energies were calculated for the atomic O adsorbed on all non-equivalent symmetric sites for the supercells with (2x2) in-plane periodicity. We have found that for all surfaces the hollow sites are preferred for O adsorption (these sites are marked with crosses in Fig. 1). The *$E_B(O)$* values for the preferred adsorption sites are listed in the first four rows of Table 1. As seen from Fig. 1, for the non-segregated alloys, oxygen prefers to make bonds with the surface Co atoms. In the case of $Pd_{0.75}Co_{0.25}$, it makes one O – Co and two O – Pd bonds with *$E_B(O)$* larger than that for pure Pd. In the case of $Pd_{0.5}Co_{0.5}$, there are two O – Co and one O – Pd bonds, which results in further strengthening of oxygen bonding to the surface. It is important to note that the oxygen – surface bonding is found to be weaker for Pd/$Pd_{0.75}Co_{0.25}$ than for pure Pd. This result supports the assumption [16, 19] that an optimal catalyst for ORR has to have lower *$E_B(O)$* than that for Pd. On the other hand, the stronger oxygen – surface bonding, found for the non-segregated alloys, suggests that these materials will not catalyze ORR efficiently.

Since, in the course of ORR, hydroxil may be formed upon "landing" of a proton and electron on the adsorbed oxygen, *$E_B(OH)$* were calculated for the preferred adsorption sites obtained for atomic O. The results are shown in the last four rows of Table 1. Note that changes in *$E_B(O)$* and *$E_B(OH)$* upon varying of the catalyst composition have the same trend, with a slight deviation from a linear relation.

### 4.2. Co-adsorption of the ORR intermediates and water

Although it is not easy to measure the coverage of the ORR intermediates in real reaction environment and it depends on electrode potential and varies from one material to other, there are indications [17, 32, 33] that, in general, the coverage is higher than that (0.25 ML) considered above. Furthermore, in the course of the reaction, the intermediates occur to be co-adsorbed with each other or with $H_2O$ at the neighboring surface sites. To study the effect of such co-adsorption we have calculated $E_B(O)$, $E_B(OH)$, and $E_B(H_2O)$ for the 0.5 ML coverage of each specie, as well as for the O – OH, O – H$_2$O, and OH – H$_2$O co-adsorption using the (2x2) supercell. These calculations were performed for Pd and Pd/Pd$_{0.75}$Co$_{0.25}$. In initial configurations, O and OH were placed at neighboring hollow sites, which are found to be most stable for the 0.25 ML coverage, while water molecule was placed at a top site. To avoid an artificial force cancelation at symmetric sites, position and orientation of the adsorbates were slightly disturbed.

The relaxed configurations obtained for Pd/Pd$_{0.75}$Co$_{0.25}$ are shown in Figs. 2 and 3 and the calculated $E_B$ values are listed in the Table 2. As seen from Fig. 2, O adsorbed with 0.5 ML coverage keeps staying at hollow sites reflecting symmetry of the system. Due to electronic charge transfer from the metal surface, oxygen atoms become negatively charged and thus repel each other. This repulsion causes an increase in the total energy of the system which leads to a significant decrease in the binding energy. Indeed, the increase in the O coverage from 0.25 ML to 0.5 ML causes the decrease in $E_B(O)$ by 0.579 eV and 0.684 eV for Pd and Pd/Pd$_{0.75}$Co$_{0.25}$, respectively (see Tables 1 and 2).

For the O – OH co-adsorption, we find that OH moves upon relaxation from the fcc hollow site to a bridge and tilts towards O adsorbed at the next fcc hollow site. It may be explained as a result of attraction between positively charged H and negatively charged O. This reordering reduces the total energy of the system. However, the overall effect is a significant weakening of both OH and O bonding to both Pd and Pd/Pd$_{0.75}$Co$_{0.25}$ caused by O – OH repulsion.

We find the hydroxyl adsorbed with 0.5 ML coverage to have two stable configurations. The first one is achieved if in an initial configuration two OH are placed in two neighboring fcc hollow sites and tilted by a few degrees (to avoid artificial force cancelation). In this case, in the course of relaxation, both molecules keep staying at these hollow sites and take positions normal to the surface. However, it appears to be a local minimum with $E_B(OH)$ equal to 1.812 eV and 1.962 eV for Pd and Pd/Pd$_{0.75}$Co$_{0.25}$, respectively. Indeed, if we initially tilt OH by ~30° and move by 0.03 Å from the symmetric position (which is achievable within the frustrated rotation vibrational mode), the adsorbates undergo restructuring upon relaxation which results in the configuration shown in the left panel of Fig. 3. This configuration is found to be much more stable than the symmetrical one. As a result we can conclude that upon increase in hydroxyl coverage from 0.25 ML to 0.5 ML the OH bonding to metal surface strengthens with increase in $E_B(OH)$ by 0.185 eV and 0.196 eV for Pd and Pd/Pd$_{0.75}$Co$_{0.25}$, respectively. We may assume that the hydrogen bonds, made between the adsorbed species for the arrangement of OH shown in the figure, lead to the decrease in the total energy of this system and hence to the increase in $E_B(OH)$.

Co-adsorption with water also affects energetics and geometrical structure of the ORR intermediates. We find that in the O – H$_2$O co-adsorbed structure the O atoms stay at the initial hollow site, while H$_2$O slightly shifts from the center of the top site and tilts to make hydrogen bonds with O. As a result, the co-adsorption causes increase in $E_B(O)$ by 0.223 eV and 0.073 eV for Pd and Pd/Pd$_{0.75}$Co$_{0.25}$, respectively.

We find that co-adsorption of hydroxyl with water also strengthens the OH bonds to the catalyst surface: $E_B(OH)$ increases upon the co-adsorption by 0.301 eV and 0.333 eV for Pd and Pd/Pd$_{0.75}$Co$_{0.25}$, respectively. Note that similar effect has been reported for other metal surfaces [16, 35]. As seen from Fig. 3, in the OH – H$_2$O structure, OH is shifted significantly from the hollow toward the bridge sites and tilted to make hydrogen bonds with H$_2$O, which stabilize the system. The water molecules are also found to be stabilized on the surfaces upon O and OH co-adsorption. For example, as compared to 0.25 ML H$_2$O adsorption, binding energy of water to Pd(111) is increased from 0.299 eV to 0.448 eV and to 0.585 eV upon O and OH co-adsorption, respectively.

### 4.3. Reaction free energy diagrams

The calculated adsorption energies, as well as zero point energies (shown in Table 1) and entropic contributions were used to build the reaction free energy diagrams. We focus on the reaction steps described by Eqs. 9 and 10. We thus include the following states in the diagrams: 1/2O$_2$ + H$_2$ gas phase, O* + H$_2$, *OH + 1/2H$_2$, and H$_2$O + *. Although these steps actually correspond to the dissociative reaction path, they also are a part of the associative path and, as mentioned in Introduction, they are likely to be the rate limiting for ORR [16]. The diagrams were constructed for Pd and Pd/Pd$_{0.75}$Co$_{0.25}$ with the 0.25 ML and 0.5 ML coverage of adsorbates, as well as for configurations with O – OH, O – H$_2$O, and OH – H$_2$O co-adsorption. Figs. 4 and 5 show the diagrams built for $U = 0$ and pH = 0.

Since 0.25 ML coverage of the ORR intermediates is used in many calculations [4,5,14,15,34] to characterize electrocatalytic activity of metal or alloy surfaces, we first built the reaction free energy diagrams for this coverage. As seen from Figs. 4 and 5, for both Pd and Pd/Pd$_{0.75}$Co$_{0.25}$, $G(O) - G(OH)$ is much smaller than $G(OH)$. Therefore, within the model that we use, the onset potential is determined by the O* + H$^+$ + e$^-$ → HO* reaction step and its value can be estimated as $U_0 = G(O) - G(OH)$. This estimate results in $U_0$ equal to 0.30 V and 0.37 V for Pd and Pd/Pd$_{0.75}$Co$_{0.25}$, respectively, which are much smaller than the experimental values. It is important to note here that this model operates with the thermodynamic quantities and does not take into account kinetic barriers. Therefore, it is expected to overestimate the $U_0$ values. The fact that our calculations result in $U_0$ smaller than in experiment thus suggests that modeling of ORR with the 0.25 ML coverage of the ORR intermediates considered above is not realistic. There are some indications that the solvent effects can change the energetics of these reaction steps [36]. However, we focus here on the coverage and co-adsorption effects. As shown in Figs. 4 and 5, the ORR energetics changes dramatically upon co-adsorption and variation of coverage of the intermediates. The co-adsorption with oxygen causes a significant increase in both $G(O)$ and $G(OH)$. However, the $G(O) - G(OH)$ difference and hence $U_0$ remain small. Water co-adsorption slightly improves the reaction energetics. Since it stabilizes the OH adsorption more than the O one, the $G(O) - G(OH)$ difference increases giving $U_0$ equal to 0.38 V and 0.63 V for Pd and Pd/Pd$_{0.75}$Co$_{0.25}$, respectively. Note that we modeled co-adsorption of O and OH with 0.25

ML of $H_2O$ which is certainty less than that in real reaction environment. One may expect that higher coverage of the co-adsorbed water can further improve the reaction energetics.

The most pronounced effect is caused by hydroxyl co-adsorption. As shown in the previous subsection, $E_B(O)$ decreases significantly upon OH co-adsorption, while the increase in the OH coverage from 0.25 ML to 0.5 ML slightly stabilizes the system. It leads to a significant increase in $G(O)$ and decrease in $G(OH)$. As a result, we obtain for both Pd and $Pd/Pd_{0.75}Co_{0.25}$ $G(O)$ to be greater than $2G(OH)$, which makes the $HO^* + H^+ + e^- \rightarrow H_2O + *$ reaction step determining for the onset potential. Applying $U_0 = G(OH)$, we find $U_0$ to be equal to 0.79 V and 0.87 V for Pd and $Pd/Pd_{0.75}Co_{0.25}$, respectively. These results are in a very good agreement with experiment [37]. This finding raises the question whether the OH co-adsorption really determines the reaction energetics, or the agreement with experiment is a result of cancelation of effects which have not been taken into account? It is clear that overall effect of the hydroxyl co-adsorption on ORR can be important if the probability for O and OH, as well as for OH and OH, to be adsorbed at neighboring sites in the course of the reaction is high. Hydroxyl can be formed as a result of several possible reaction steps: a) $O^* + H^+ + e^- \rightarrow HO^*$, b) $*OOH + * \rightarrow O^* + HO^*$, c) $HOOH^* + * \rightarrow 2HO^*$ [4]. If water is involved in reaction implicitly it also produces OH at some steps [17]. It was also shown that for OH co-adsorbed with water in an $O_2$ free environment on Pt(111) the OH – OH interaction is attractive up to the 1/3 ML coverage of OH [17, 32]. Based on this consideration, one may expect that in real reaction environment, in which all adsorbate configurations corresponding to various reaction steps are present, probability of the O – OH and OH – OH co-adsorption is high and its effect in the ORR energetics can be substantial.

We find the OH co-adsorption to be favorable for the reaction energetic. However, it is known [1, 38] that increase in the OH coverage can reduce the ORR rate by blocking active sites for the $O_2$ adsorption. These two (thermodynamic and kinetic) effects may be competing. On the other hand, as discussed above, the O – OH and OH – OH co-adsorption can be achieved without significant increase in the OH coverage.

**4.4. Tracing the reaction energetics to the electronic structure of the alloys**

As shown above, the binding energies of the ORR intermediates and hence the reaction free energies change significantly upon variation of composition of the Pd-Co alloys. Since chemisorption is determined by hybridization between the electronic states of adsorbate and surface atoms, in this subsection, we evaluate the effect of the surface composition on the hybridization by analyzing LDOS of the adsorbed oxygen and surface atoms for the alloys under consideration. As a representative example, in Fig. 6, we show LDOS of the *d*Pd- and *d*Co-states of the surface atoms and *p*O-states of the adsorbed oxygen calculated for the $Pd_{0.75}Co_{0.25}$ alloy. One can see that the hybridization of the *d*Pd- and *p*O-states with the spin-polarized *d*Co states induces spin-polarization for the formers. The non-occupied states of O, which are important for chemisorptions, are formed in the system due to hybridization with the spin-down *d*Co-states.

Since the initial $d$Pd states are mostly overlapped energetically with the spin-up $d$Co-states, their hybridization determines the alloying effect on the Pd LDOS. As we shall see, this effect leads to a low-energy shift of the Pd $d$-band.

Since the spin-polarized LDOS have a quite complicated structure, to evaluate the overall effect of hybridization on the chemisorptions of oxygen, we analyze the summed spin-up and spin-down LDOS. As we have shown above the Pd/Pd$_{0.75}$Co$_{0.25}$ structure has the most promising reaction energetic among the alloys under consideration. Therefore, we compare and contrast the summed $d$Pd and $p$O LDOS for this alloy surface and clean Pd surface (as a reference), both adsorbed with oxygen (see Figs. 7 and 8). One can see two distinguished peaks (A and B) formed in the $p$O LDOS which align with two $d$Pd LDOS peaks. This suggests a significant $p$O – $d$Pd hybridization resulting in formation of anti-bonding and bonding states represented by A and B peaks, respectively. It is known that the lesser anti-bonding states are populated, the stronger covalent bonding is. Taking into account that the A peak in both systems is almost totally depopulated, we can thus use the ratio of the A peak intensity to the B peak intensity as a qualitative descriptor of the strength the $p$O – $d$Pd covalent bonding in the systems. We find this ratio to be equal to 0.35 and 0.29 for Pd and Pd/Pd$_{0.75}$Co$_{0.25}$, respectively, which leads us to the conclusion that adsorbed oxygen makes stronger covalent bonds to Pd than to Pd/Pd$_{0.75}$Co$_{0.25}$.

The next question to rise is why the O – Pd covalent bonding is stronger on pure Pd than on the Pd/Pd$_{0.75}$Co$_{0.25}$ surface. To answer this question we analyze LDOS of surfaces atoms of these systems without adsorbate. LDOS of surface Pd atoms calculated for Pd(111) and Pd/Pd$_{0.75}$Co$_{0.25}$ (111) are plotted in Fig. 9. Note that for Pd/Pd$_{0.75}$Co$_{0.25}$ there are two kinds of Pd surface atoms: one has no Co neighboring atoms and the other has one Co neighbor. LDOS of the latter is shown in Fig. 9. One can see that the density of the surface Pd $d$-states around the Fermi-level for Pd/Pd$_{0.75}$Co$_{0.25}$ is significantly reduced and entire $d$-band is shifted towards lower energies compared to those for Pd(111). As seen from Fig. 9, these effects are more pronounced for Pd$_{0.75}$Co$_{0.25}$ and even more for Pd$_{0.5}$Co$_{0.5}$. It is important to note that the surface Pd atom in Pd(111) naturally has no Co neighbors, while in Pd/Pd$_{0.75}$Co$_{0.25}$, Pd$_{0.75}$Co$_{0.25}$, and Pd$_{0.5}$Co$_{0.5}$ it has one, three, and six nearest Co neighbors, respectively. We thus find a close correlation between energetic position of the $d$-band of surface Pd atom and the number of its Co-nearest neighbors: the more Co neighbors Pd atom has the deeper its $d$-band is located. This correlation reflects the discussed above effect of the $d$Pd – $d$Co hybridization, which causes a low energy shift of the $d$Pd band. Position of the Pd $d$-band center with respect to the Fermi-level, plotted versus number of Pd-Co bonds, clearly illustrates this correlation (see Fig. 10). This finding is very important, because the quantity of our interest – strength of covalent O – Pd bonding is determined by hybridization of both occupied and non-occupied oxygen and metal states, which in turn depends on the density of the metal $d$-states around the Fermi-level. If the $d$-band shifts towards lower energies upon composition variation, the density of $d$-states around the Fermi-level decreases which causes weakening the oxygen metal bond. This effect is depicted in the simple and widely used model [39] which correlates the $d$-band center position with the adsorption energy of oxygen or other species. We can thus conclude from our results that the hybridization of $d$Pd – $d$Co states in Pd/Pd$_{0.75}$Co$_{0.25}$ causes a low-energy shift of the $d$-band center of the surface Pd atoms, with respect to that of the Pd(111), which in turn leads to a weakening of the O – Pd covalent bond and a decrease in the O binding energy. Note that, as we move from Pd and Pd/Pd$_{0.75}$Co$_{0.25}$ to Pd$_{0.75}$Co$_{0.25}$ and Pd$_{0.5}$Co$_{0.5}$, the Pd $d$-band center is shifted toward lower energies, while the O bonding is strengthening significantly. This happens because Pd$_{0.75}$Co$_{0.25}$ and Pd$_{0.5}$Co$_{0.5}$ have surface Co atoms and the O binding energy is determined rather by stronger Co – O bonding.

As mentioned in Introduction, the authors of Ref. 14 propose the contraction of Pd – Pd bonds caused by alloying with Co to be the factor which changes the electronic structure and binding energy of oxygen. Our results bring us to a different conclusion. As seen from Figs. 9 and 10, the Pd $d$-band center is deeper for Pd$_{0.75}$Co$_{0.25}$ than for Pd/Pd$_{0.75}$Co$_{0.25}$ even though these two systems have the same lattice parameter and hence the same Pd – Pd bond length. Furthermore, as shown in Fig. 11, two non-equivalent Pd surface atoms of Pd/Pd$_{0.75}$Co$_{0.25}$, have significantly different LDOS: the Pd atom that has a Co nearest neighbor has lower density of states around the Fermi-level and deeper the $d$-band center than the other Pd surface atom that has only Pd nearest neighbors. Our results thus bring us to the conclusion that the $d$Pd – $d$Co hybridization is the main factor which controls LDOS of Pd and the oxygen bindingn energy in Pd – Co alloys with surface segregation.

## 5. Conclusions

We have carried out the DFT based computational studies of electronic structure of Pd – Co alloys, as well as energetics of adsorption, and vibrational frequencies of the ORR intermediates on the alloy surfaces. We find the surface segregation of the alloys observed in experiment is essential for improving electrocatalytic properties of these materials. Indeed, the binding energy of the ORR intermediates on Pd/Pd$_{0.75}$Co$_{0.25}$ are found to be lower than that on Pd(111) which is favorable for the reaction, while the Pd$_{0.75}$Co$_{0.25}$ and Pd$_{0.5}$Co$_{0.5}$ surfaces are too reactive for ORR due to bonding to the surface Co atoms. Our results show that the hybridization between $d$Pd- and $d$Co-states causes the low energy shift of the $d$-band of surface Pd in Pd/Pd$_{0.75}$Co$_{0.25}$, which causes weakening the bonding of the intermediates to the surface.

Co-adsorption of the ORR intermediate and water is found to change the reaction energetics significantly. We have built the ORR free energy diagrams for the Pd and Pd/Pd$_{0.75}$Co$_{0.25}$, and estimated the onset electrode potential for the reaction. For the intermediates adsorbed with 0.25 ML coverage, the estimated from the calculations $U_0$ is found to be much lower than in experiment. Co-adsorption with water slightly improves the results, while the diagrams built for the O – OH and OH – OH co-adsorption configurations provide $U_0$ which are in a good agreement with experiment. The oxygen binding energies, obtained for Pd and Pd/Pd$_{0.75}$Co$_{0.25}$ support the assumption [21,22] that $E_b(O)$ can be used as a descriptor of the ORR activity.

Table 1. Binding energies and zero point vibration energies calculated for O and OH adsorbed with 0.25 ML coverage on the preferred sites of the (111) surfaces of Pd and Pd-Co alloys.

| Adsorbate | Slab | $E_B$ (eV) | ZPE (eV) |
|---|---|---|---|
| O | Pd | 4.568 | 0.0695 |
| | $Pd_{0.75}Co_{0.25}$ | 4.930 | 0.0626 |
| | $Pd_{0.50}Co_{0.50}$ | 5.620 | 0.0634 |
| | Pd/ $Pd_{0.75}Co_{0.25}$ | 4.414 | 0.0690 |
| OH | Pd | 2.608 | 0.3309 |
| | $Pd_{0.75}Co_{0.25}$ | 3.032 | 0.3318 |
| | $Pd_{0.50}Co_{0.50}$ | 3.464 | 0.3418 |
| | Pd/ $Pd_{0.75}Co_{0.25}$ | 2.522 | 0.3310 |

Table 2. Binding energies of O and OH calculated for the co-adsorption configurations.

| Slab | Co-adsorbate | $E_B(O)$, eV | $E_B(OH)$, eV |
|---|---|---|---|
| Pd | O | 3.989 | 2.004 |
| | OH | 4.038 | 2.793 |
| | $H_2O$ | 4.791 | 2.909 |
| Pd/Pd $_{0.75}Co_{0.25}$ | O | 3.730 | 2.068 |
| | OH | 4.038 | 2.718 |
| | $H_2O$ | 4.487 | 2.855 |

**Figure Captions**

Fig. 1. (Color online). Top view of the surfaces studied in this work: a) Pd(111), b) $Pd_{0.25}Co_{0.75}$, c) $Pd_{0.5}Co_{0.5}$, and d) $Pd/Pd_{0.25}Co_{0.75}$. Light grey and dark blue balls represent the Pd and Co atoms, respectively. Black strait lines separate supercells. Black crosses mark the preferred adsorption sites for atomic oxygen.

Fig. 2. (Color online). The lowest energy configurations for the 0.5 ML oxygen adsorption (left panel) and O – OH co-adsorption (right panel). Medium dark red and small light blue balls represent the O and H atoms, respectively.

Fig. 3. (Color online). The lowest energy configurations for the 0.5 ML hydroxyl adsorption (left panel) and OH – $H_2O$ co-adsorption (right panel).

Fig. 4. (Color online). Reaction free energy diagram built for Pd for the following intermediate adsorption configurations: 0.25 ML coverage of the adsorbates (thick solid lines), O – OH and OH – OH co-adsorption (thin solid lines), O – O and O – OH co-adsorption (dash lines), and O – $H_2O$ and OH – $H_2O$ co-adsorption (dash-dot lines). The arrowed line shows the $U_0$ determining reaction step.

Fig. 5. (Color online). Reaction free energy diagram built for $Pd/Pd_{0.25}Co_{0.75}$. Adsorption configurations and line code is the same as in Fig. 4.

Fig. 6. (Color online). Spin-resolved LDOS of the surface atoms and adsorbed oxygen calculated for $Pd_{0.25}Co_{0.75}$.

Fig. 7. (Color online). Spin-summed LDOS of the surface Pd atom and adsorbed oxygen calculated for Pd(111).

Fig. 8. (Color online). Spin-summed LDOS of the surface Pd atom and adsorbed oxygen calculated for $Pd/Pd_{0.25}Co_{0.75}(111)$.

Fig. 9. (Color online). Spin-summed LDOS of the (111) surface Pd atom calculated for pure Pd (solid line), $Pd/Pd_{0.25}Co_{0.75}$ (dash line), $Pd_{0.25}Co_{0.75}$ (dash-dot-dot line), and $Pd_{0.5}Co_{0.5}$ (dot line).

Fig. 10. Energy of the Pd *d*-band center counted fro the Fermi-level as a function of the neighboring Co atoms.

Fig. 11. (Color online). Spin-summed Pd LDOS calculated for $Pd/Pd_{0.25}Co_{0.75}(111)$ for the surface Pd atoms which have no Co neighbor (solid line) and has one Co neighbor (dash line).

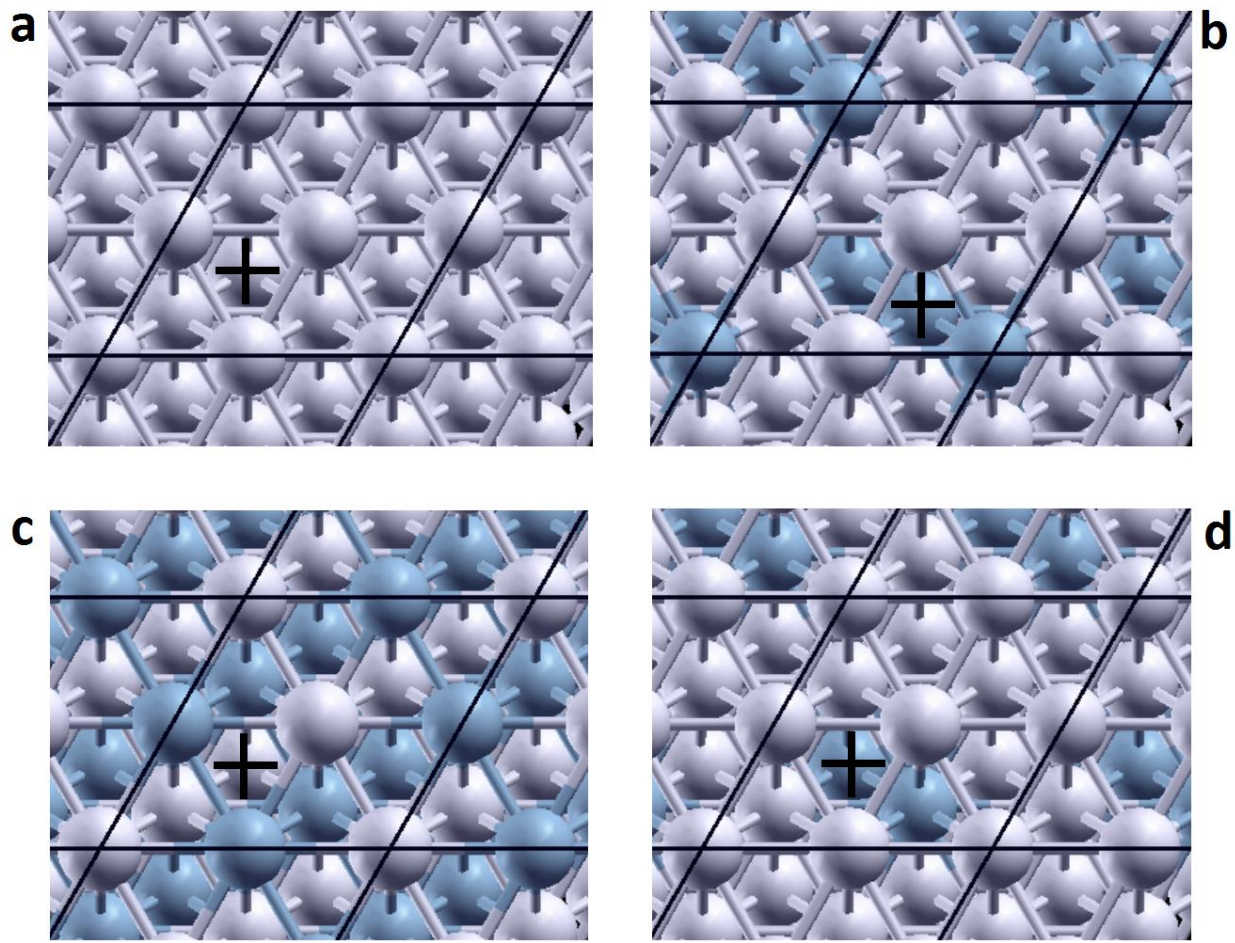

Fig. 1.

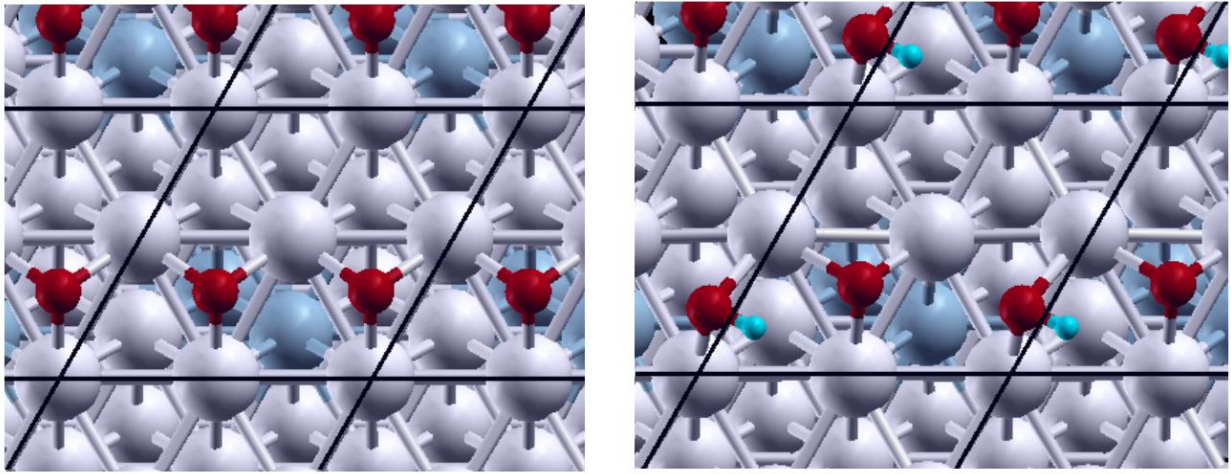

Fig. 2.

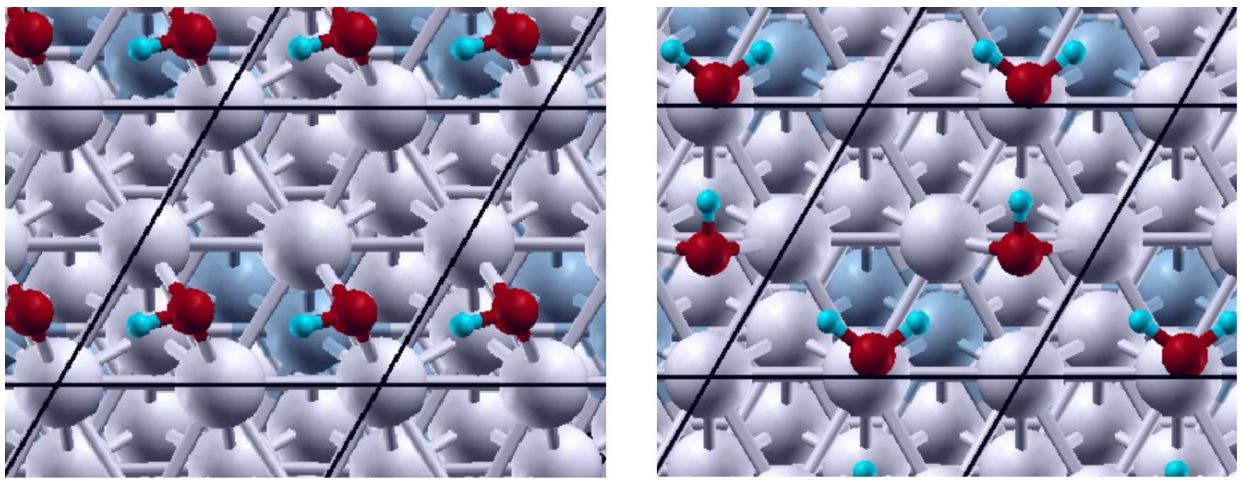

Fig. 3.

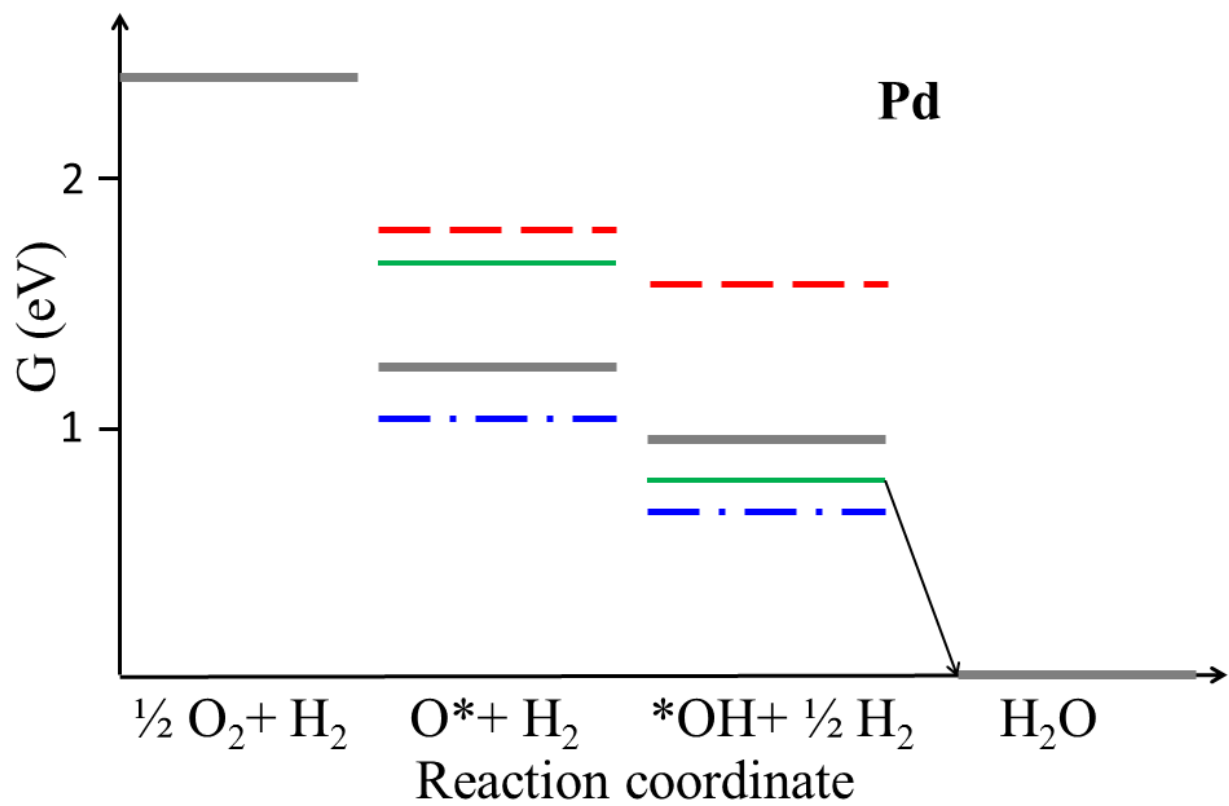

Fig. 4.

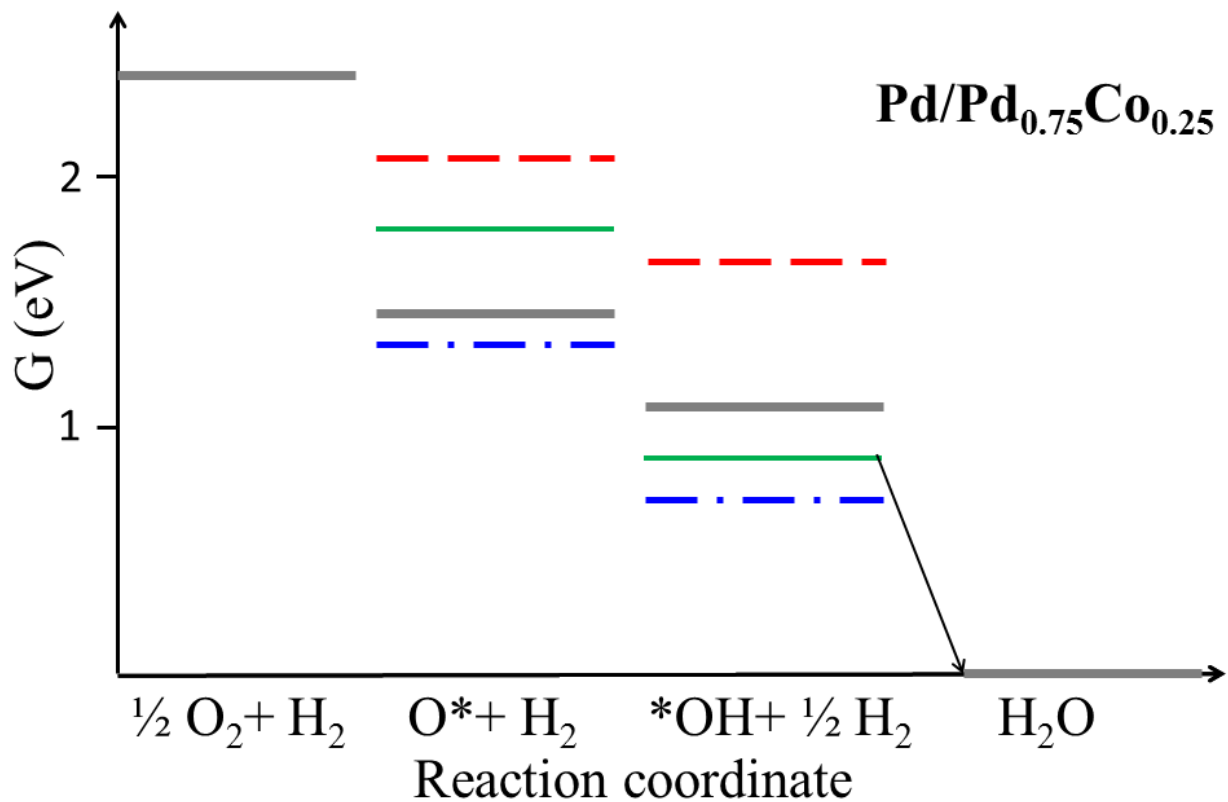

Fig. 5.

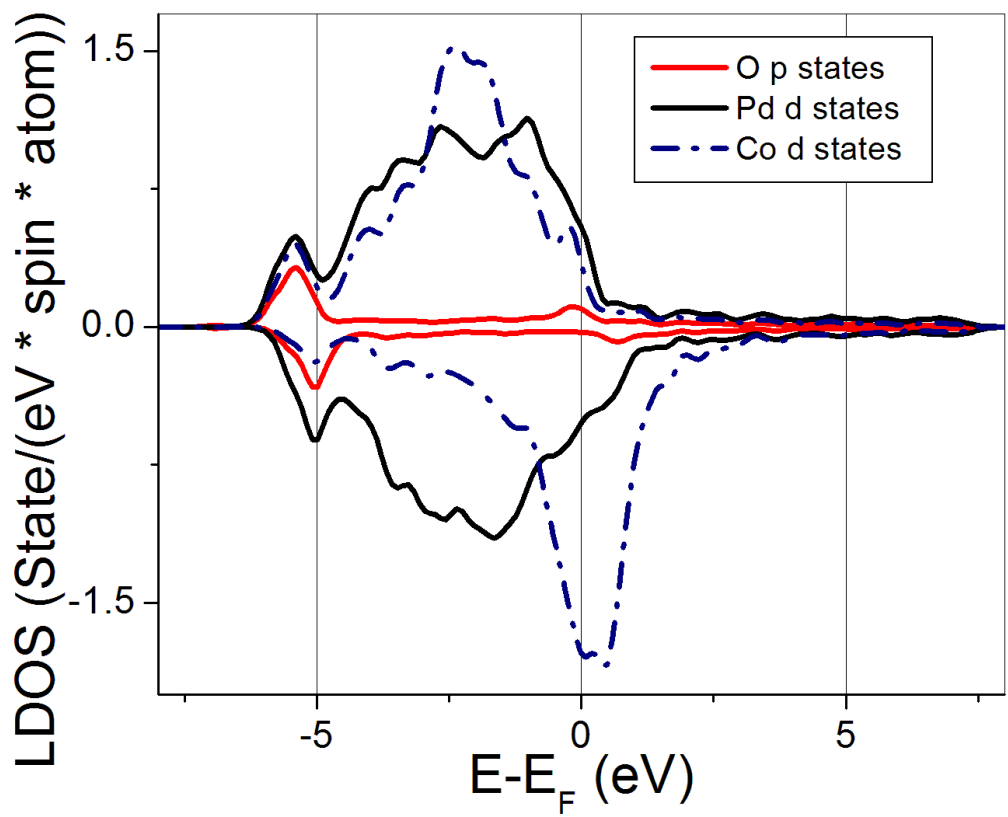

Fig. 6.

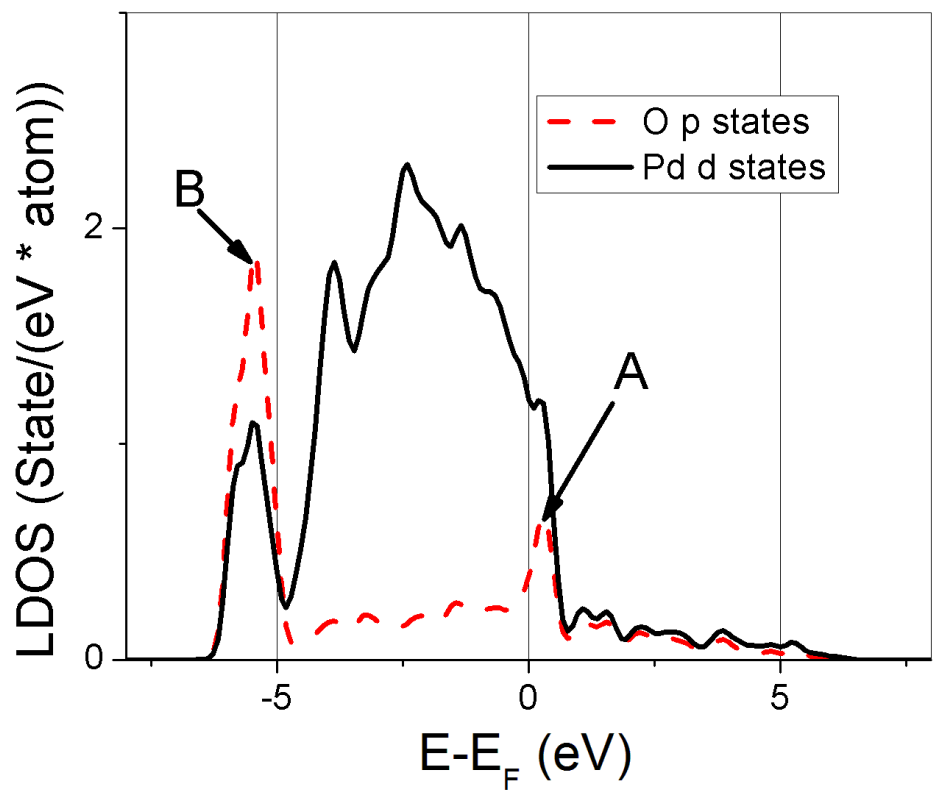

Fig. 7.

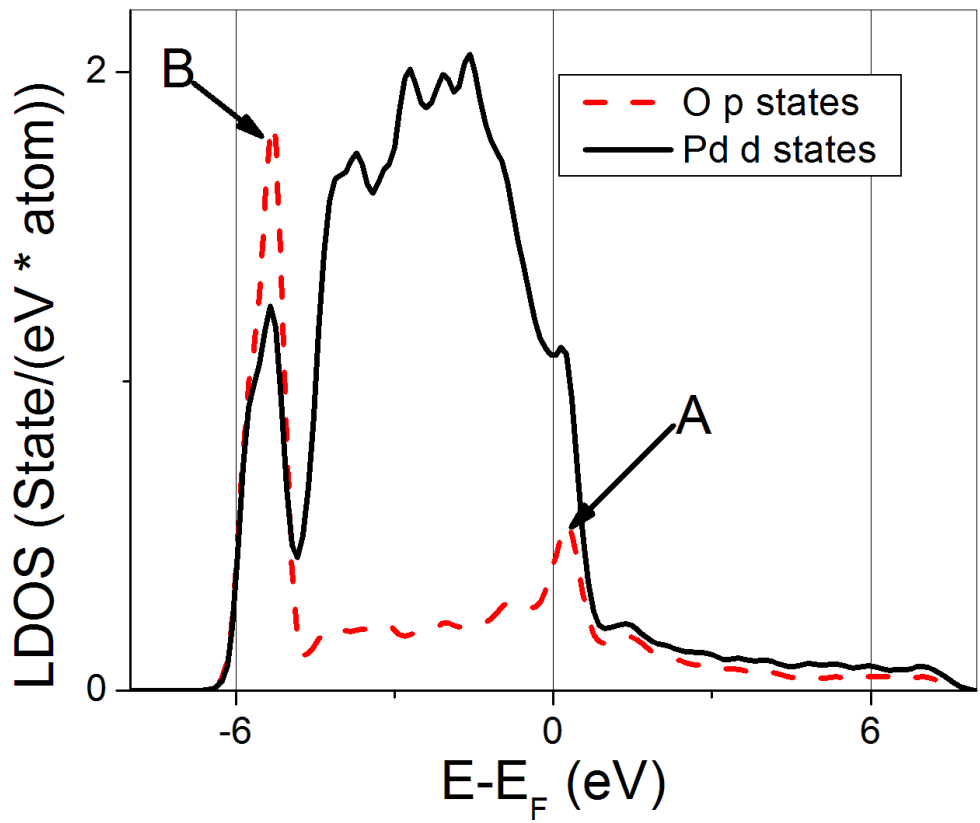

Fig. 8.

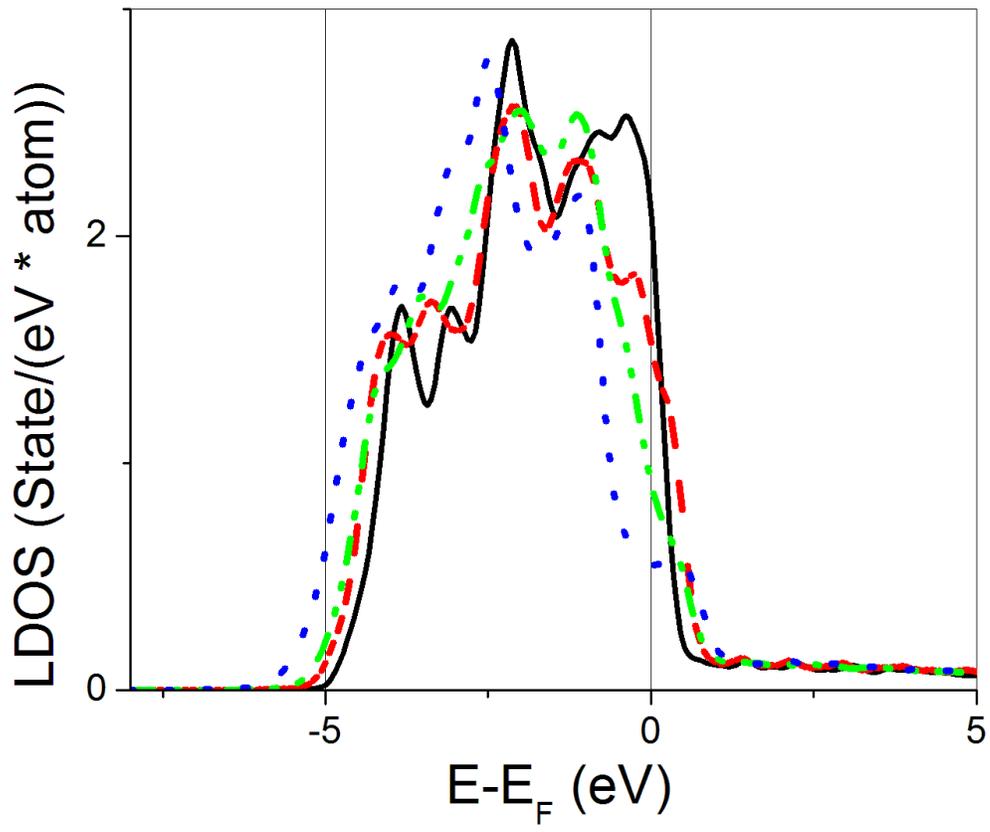

Fig. 9.

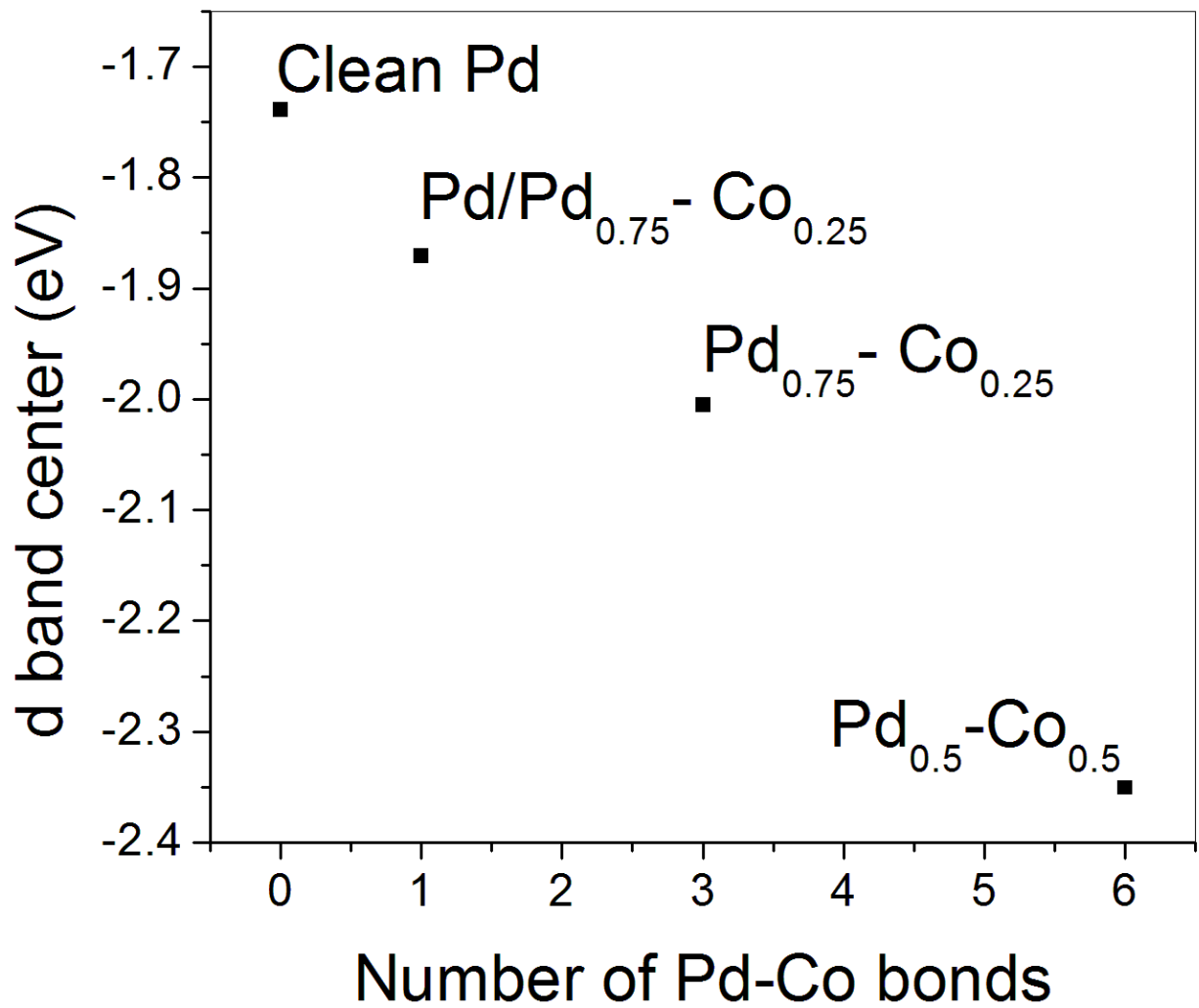

Fig. 10.

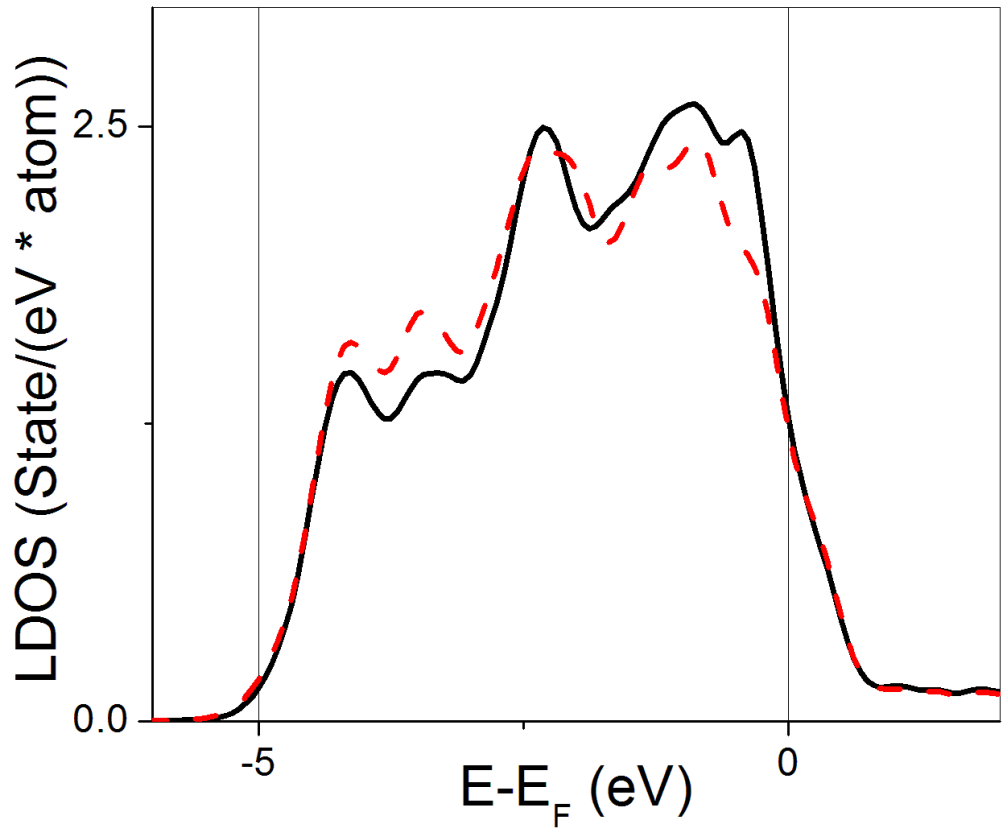

Fig. 11.